\documentclass[twocolumn,aps,prl,floatfix,superscriptaddress]{revtex4-2}

\usepackage{hyperref}
\usepackage[utf8]{inputenc}

\RequirePackage{lineno}

\usepackage{graphicx}
\usepackage{dcolumn}
\usepackage{amsmath}
\usepackage{supertabular}
\usepackage{multirow}
\usepackage{color}
\usepackage{cancel}
\usepackage{ulem}
\usepackage{array}
\usepackage{tabularx}
\usepackage{orcidlink}
\usepackage{hyperref}

\usepackage{amsmath}
\usepackage{amssymb}
\usepackage{xcolor}
\definecolor{myOrange}{rgb}{1,0.5,0.}
\definecolor{myGreen}{rgb}{0.0,0.6,0.1}
 
\newcommand{\removetext}[1]{}

\newcommand{\sqrts}{$\sqrt{s}$}
\newcommand{\pp}{$pp$}
\newcommand{\epA}{$ep/A$}
\newcommand{\JetMass}{$M_{\rm jet}$}
\newcommand{\JetEta}{$\eta_{\rm jet}$}
\newcommand{\JetPt}{$p_{\rm T,jet}$}
\newcommand{\JetPhi}{$\phi_{\rm jet}$}
\newcommand{\JetCharge}{$Q^{\rm ch}_{\kappa}$}
\newcommand{\girth}{$girth$}
\newcommand{\qqbar}{$q-\bar{q}$}
\newcommand{\JetNConst}{$N_{\rm const}$}
\newcommand{\gevc}{$\rm GeV/{\it c}$}
\newcommand{\gammaJet}{$\gamma + {\rm jet}$}
 
\begin{document}


\title{Discriminating QCD Compton and Quark–Antiquark Annihilation Processes in $\gamma$+Jets Using Interpretable Machine Learning}

\bigskip

\author{Monalini Samal}
\affiliation{Indian Institute of Science Education and Research (IISER), Berhampur 760010, India}

\author{Nihar Ranjan Sahoo
\orcidlink{0000-0003-4518-6630}
}
    \affiliation{Indian Institute of Science Education and Research (IISER), Tirupati 517619, India}
\date{\today} 


\begin{abstract}
We investigate how effectively final-state jet substructure can discriminate between QCD Compton and quark--antiquark annihilation processes from photon--jet production in \pp\ collisions at $\sqrt{s}=13$ TeV. Using infrared- and collinear-safe jet observables, multivariate classifiers---boosted decision trees and multilayer perceptrons---are trained on labeled quark- and gluon-initiated jets from dijet events and applied to photon--jet samples. Observables probing soft and wide-angle radiation, in particular jet multiplicity and jet girth, dominate the discrimination. The jet mass provides a complementary but weaker contribution, while the jet charge exhibits negligible discriminating power. A comparison of the two classifiers demonstrates that the achievable separation is limited primarily by QCD radiation effects rather than by classifier complexity. These findings quantify the extent to which information about the underlying hard process survives hadronization and realistic jet reconstruction, providing a physics-driven baseline for precision jet measurements in \pp, \epA, and heavy-ion collisions.
\end{abstract}

\maketitle

\section{Introduction}

Quark–gluon jet discrimination is a long-standing problem in Quantum Chromodynamics (QCD), rooted in the different color representations of quarks ($C_{F}=4/3$) and gluons ($C_{A}=3$) and their associated radiation patterns~\cite{AMY:1989rdg,OPAL:1991ssr}. Due to their larger color factor, gluons radiate more copiously than quarks, leading on average to jets with broader energy distributions, larger particle multiplicities, and enhanced soft radiation. These differences motivate the use of jet substructure observables as experimental proxies for the identity of the initiating parton 
~\cite{Dokshitzer:1991wu,Larkoski:2019nwj,Metodiev:2018ftz}. However, the distinction between quark- and gluon-initiated jets is intrinsically probabilistic and degrades at high energies due to multiple emissions, color coherence, and Sudakov suppression~\cite{Gallicchio:2012ez,Larkoski:2017jix}.\\

Photon-tagged jet (\gammaJet) measurements provide a particularly clean environment, compared to inclusive jet and dijet measurements, to investigate this problem. At leading order, direct photon production in \pp\ collisions proceeds dominantly through QCD Compton scattering $(qg \rightarrow q\gamma)$ and quark–antiquark annihilation $(q\bar{q} \rightarrow g\gamma)$, resulting in recoil jets that are preferentially quark- or gluon-initiated, respectively. While the underlying hard process is well defined at the parton level, it remains a nontrivial question how much of this information survives parton showering, hadronization, and realistic jet reconstruction.\\


Understanding this distinction is not only of conceptual importance but also directly relevant for precision QCD studies and jet measurements in both \pp\ and heavy-ion collisions. In heavy-ion environments, where jets serve as probes of the quark–gluon plasma (QGP), disentangling quark- and gluon-dominated jet samples is essential for interpreting color-charge–dependent energy loss and medium-induced modifications ~\cite{STAR:2023pal,STAR:2023ksv,STAR:2025yhg,ALICE:2023qve}. Even in proton–proton collisions, quantifying the intrinsic limitations of process-level discrimination based on final-state observables provides an important baseline for experimental analyses.\\

Machine-learning (ML) techniques---such as boosted decision trees (BDT) and multilayer perceptrons (MLP)---provide a systematic experimental approach to quark–gluon discrimination by combining multiple jet substructure observables~\cite{Gallicchio:2012ez,Baldi:2014} and exploiting their correlations. These methods act as efficient estimators of the information contained in the jet substructures. Comparing different multivariate classifiers allows us to assess how much discriminating power can be extracted from experimentally accessible observables and to identify where performance saturates due to intrinsic QCD effects such as color coherence and Sudakov radiation. In this way, machine learning serves not only as a practical analysis tool but also as a probe of the fundamental limits of quark–gluon discrimination.\\

In this work, we explore the QCD problem discriminating quark-gluon jets by studying the discrimination of QCD Compton and quark–antiquark annihilation processes in \gammaJet\ events in \pp\ collisions at \sqrts=13 TeV. We train multivariate classifiers--MLP and BDT---on labeled quark- and gluon-initiated jets from dijet events and apply them to \gammaJet\ samples, using a compact set of infrared- and collinear-safe jet observables. Using MLP and BDT with individual observables, we quantify both the achievable discrimination and its physical origin. \\

This paper is organized as follows. Section~\ref{sec:Methodology} describes the methodology, including the jet observables and multivariate analysis techniques. Section~\ref{sec:Results} presents the results and discussion, and Section~\ref{sec:SummaryOutlook} summarizes the main conclusions and future prospects.\\
\begin{figure*}
    \centering
    \includegraphics[width=1\textwidth]{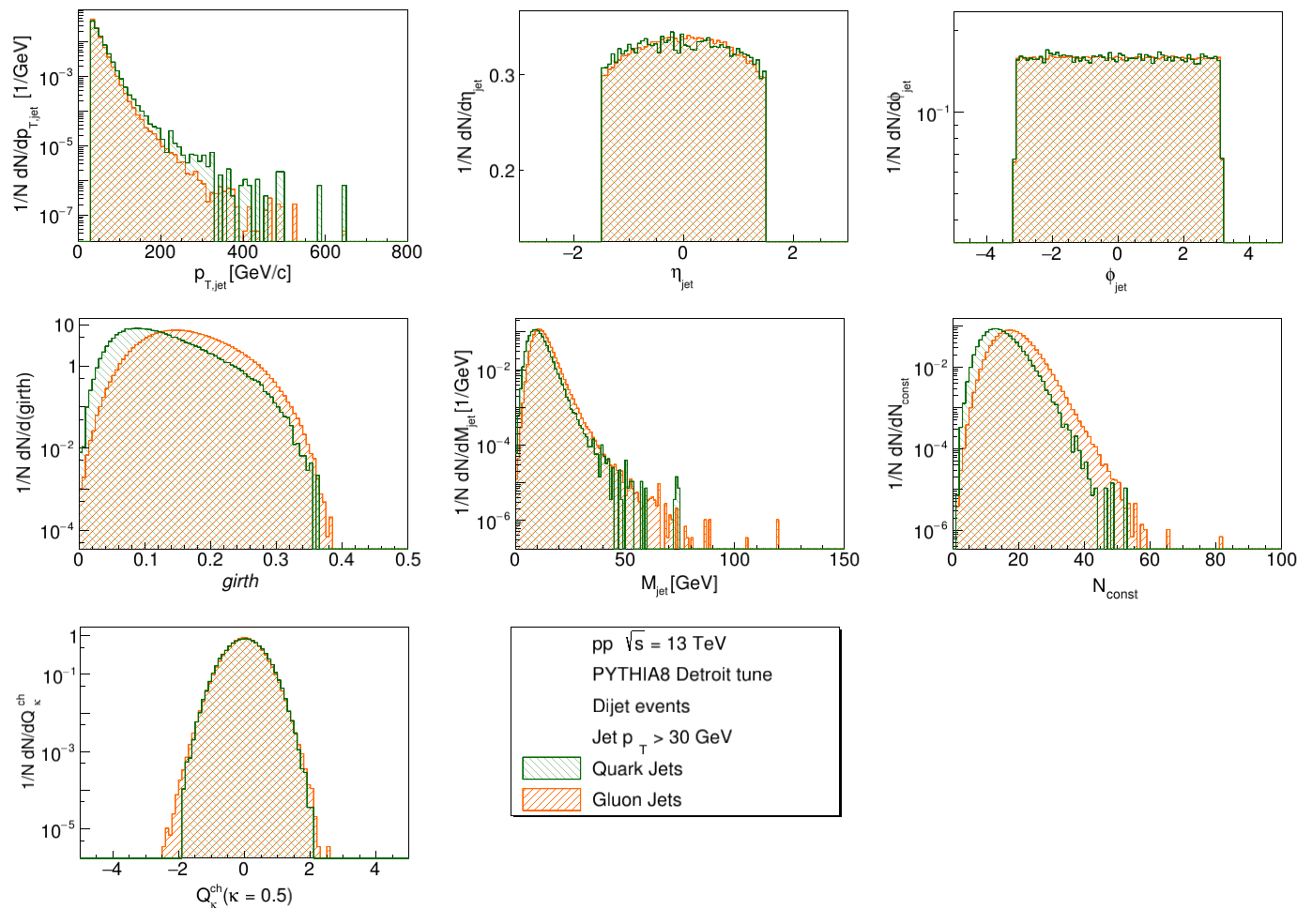}
    \caption{Seven input features for Quark (green) vs. Gluon (orange) Jet used for training: \JetPt, \JetEta, \JetPhi, \JetNConst, \JetMass, \JetCharge. These quantities are calculated from \pp\ collisions at \sqrts=13 TeV using PYTHIA8 Detroit tune dijet sample with \JetPt $>$ 30 \gevc.}
    \label{fig:TrainingQvsG}
\end{figure*}

\section{Methodology}
\label{sec:Methodology}
\subsection{Event generation and Jet reconstruction}

Monte Carlo \pp\ events at \sqrts\ = 13 TeV are generated using the PYTHIA8-Detroit tune~\cite{Aguilar:2021sfa} with $\hat{p}_{\rm T} >$ 30 \gevc. Two classes of events are produced: inclusive dijet events and \gammaJet\ events.\\ 

Jets are reconstructed from final-state particles using anti-$k_{\rm T}$ jet-clustering algorithm provided in FastJet package~\cite{Cacciari:2008gp}.
Jet resolution parameter $R=0.5$ is used, which provides a reasonable balance between retaining jet radiation and minimizing sensitivity to soft contamination; a systematic study of the jet-radius dependence is left for future work. 
Jet flavor labels are assigned based on the flavor of the highest-$p_T$ outgoing parton matched with to the reconstructed jet from different processes. Here the matching criterion is $\Delta R <$0.25; the $\Delta R$ is the separation between the hard parton and jet axis in $\eta-\phi$ plane.
The \gammaJet\ samples are used to study the discrimination between QCD Compton and quark–antiquark (\qqbar) annihilation processes based on the flavor of the recoiling jet. Parton showering, hadronization, and underlying-event effects are modeled using the default PYTHIA8-Detroit tune. All results presented in this work are based on simulations generated with PYTHIA8-Detroit tune; no attempt is made to assess tune-dependent variations or other MC event generator, which are left for future studies.
\\

The same jet reconstruction procedure and parameters are applied consistently to both the dijet samples used for training and the \gammaJet\ samples used for process discrimination. This ensures that differences in classifier performance arise from the underlying jet structure rather than from reconstruction-related effects. For a given jet flavor, the jet substructure observables show consistent distributions between dijet and $\gamma+$jet events, indicating that the features learned from dijet samples are transferable to $\gamma+$jet recoil jets (Appendix~\ref{Appn:DiGammaJet}).

\subsection{Jet Observables}
\label{sec:JetObse}
In this multivariate analysis, jet transeverse momentum(\JetPt),  pseudorapidity (\JetEta), and azimuthal angle (\JetPhi) are used along with the jet substructure observables as input features to the BDT and MLP model. 

Although more substructure observables can be considered for this analysis, we have considered the following jet substructure observables are used:

\begin{itemize}

    \item {\bf Jet Mass:}
\begin{equation}
M_{\rm jet} \;=\; \sqrt{E_{\rm jet}^2 - p_{\rm jet}^{2}}\, .
\end{equation}

Here $E_{\rm jet}$ and $p_{\rm jet}$ are the energy and total momentum of the jet. The \JetMass\ characterizes the internal radiation structure of a jet, encoding information about the amount and angular distribution of partonic radiation contained within the jet~\cite{STAR:2021lvw,ATLAS:2012am}.

    \item {\bf Jet multiplicity:} $N_{\rm const}$ represents number of constituents in jet~\cite{ATLAS:2016vxz}. The average multiplicity in jet are also related to the Color factor, $<N_{\rm g}>/<N_{\rm q}> = C_{\rm A}/C_{\rm F}$~\cite{Gallicchio:2012ez}. 
    
    \item {\bf girth:}
\begin{equation}
girth \;\equiv\; \sum_{i \in \text{jet}} \frac{p_{T,i}}{p_{T,\text{jet}}}\,\Delta R_{i,\text{jet}},
\end{equation}
where,
\begin{equation}
\Delta R_{i,\text{jet}} \;=\; \sqrt{(\eta_i-\eta_{\text{jet}})^2+(\phi_i-\phi_{\text{jet}})^2}\,
\end{equation}
and $p_{T,i}, \eta_{i}$ and $\phi_{i}$ are the constituent's transverse momentum, pseudorapidity and azimuthal angle with respect to beam axis, respectively. The girth quantifies the radial distribution of transverse momentum within a jet and provides a measure of how broadly the jet’s energy is spread around its axis~\cite{CMS:2024zjn}.

    \item {\bf Jet Charge:} 
    \begin{equation}
Q_{\kappa}^{\rm ch} \;=\; \frac{1}{\left(p_{T,\text{jet}}^{\rm ch}\right)^{\kappa}}
\sum_{i\in \text{ch jet}} q_i \left(p_{T,i}\right)^{\kappa}.
\end{equation}
Here $q_{i}$ represents electric charge of the constituent and $\kappa$ = 0.5 that controls the weight of soft vs hard constituents~\cite{CMS:2017yer}. 

\end{itemize}

Figure~\ref{fig:TrainingQvsG} shows the distributions of the key input observables—
\JetPt, \JetMass, \JetCharge, \JetNConst, and \girth—
used to train the classifier to distinguish quark- and gluon-initiated jets in dijet events.
Jets are selected with \JetPt $>$ 30~\gevc, and the jet substructure observables
\JetMass, \JetCharge, \JetNConst, and \girth\ are computed for these jets.
These observables probe complementary aspects of the jet radiation pattern,
like the overall QCD radiation, and radial energy distribution.
While the distributions exhibit substantial overlap, systematic differences
between quark- and gluon-initiated jets are observed, providing discrimination power on a statistical basis. The overlap between the quark and gluon distributions reflects the intrinsic
limitations of quark–gluon discrimination arising from QCD radiation,motivating the use of multivariate techniques to exploit correlations among these observables.

\subsection{Multivariate Classification Models}
\label{sec:ML}
In this study, we employ two widely used and well-established multivariate techniques: a multilayer perceptron and a boosted decision tree. While more complex machine-learning architectures have been developed in recent years, our focus is on using physically transparent and experimentally robust classifiers to discriminate between QCD Compton and $q\bar{q}$ annihilation processes. These models provide a controlled framework to combine a limited set of jet substructure observables. Hence it allows us to assess how much discriminating power can be extracted from experimentally accessible inputs and to identify where the performance saturates due to intrinsic QCD effects. The input observables have clear QCD interpretations, enabling the contribution of each feature to the classifier performance to be assessed in a transparent and physically meaningful way.

\begin{figure}
    \centering
    \includegraphics[width=0.9\linewidth]{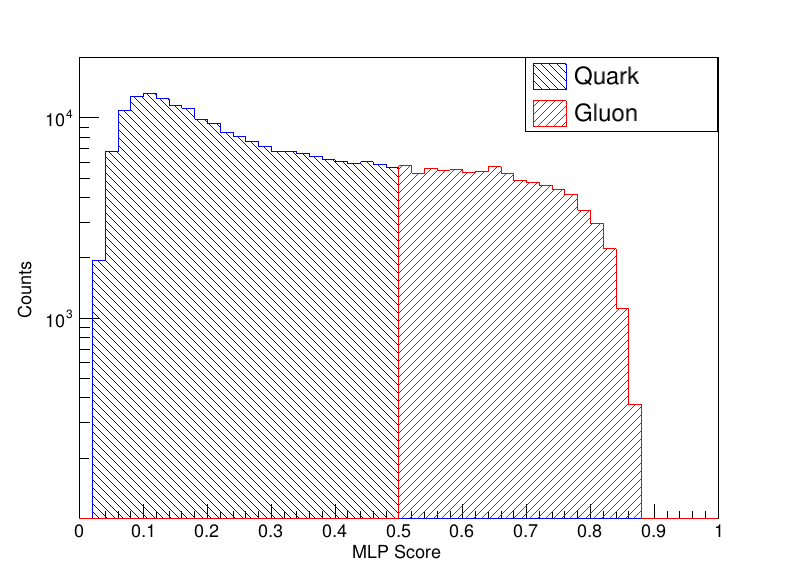}
    \caption{The MLP score distribution from \gammaJet\ sample. The MLP Score $>0.5$ ($<0.5$) is used for gluon-like (quark-like) jet. }
    \label{fig:MLPScoreQvsG}
\end{figure}

 \begin{figure}
    \centering
    \includegraphics[width=0.9\linewidth]{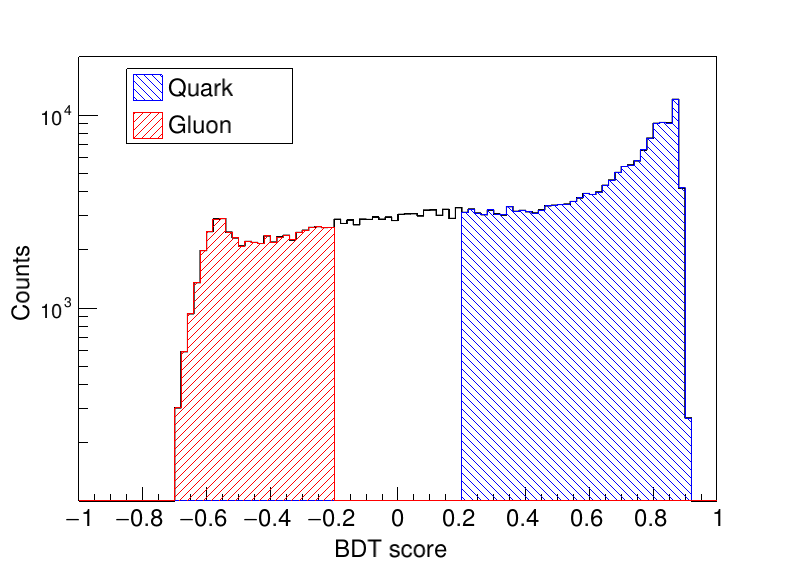}
    \caption{The BDT score distribution from \gammaJet\ sample. The BDT Score $>0.2$ ($< -0.2$) is used for quark-like (gluon-like) jet.}
    \label{fig:BDTScoreQvsG}
\end{figure}
\subsubsection{Multilayer Perceptron (MLP)}
\label{sec:MLP}
Various neural-network-based classifiers have been widely used in collider physics to capture nonlinear correlations among observables~\cite{Baldi:2014}. In this analysis, we have used a simple multilayer perceptron (MLP) classifier to study the extent to which non-linear correlations among jet substructure observables. 

The MLP classifier is implemented using the ${\it TensorFlow}$~\cite{Abadi:2016} framework with the high-level $Keras$ application programming interface (API)~\cite{Chollet:2015}. The network is trained in a supervised manner using labeled samples of dijet events in which the jets are identified at the parton level as quark- or gluon-initiated. These samples are used to learn the characteristic differences in jet substructure associated with quark and gluon radiation.\\


The trained classifier is subsequently applied to photon-tagged events to discriminate between QCD Compton and \qqbar annihilation processes, based on the flavor of the recoiling jet. \\

In this analysis, the MLP consists of seven input nodes
corresponding to the jet observables as mentioned in ~\ref{sec:JetObse}.
\begin{equation}
      {\bf x} = (p_{\rm T, jet}, \eta_{\rm jet}, \phi_{\rm jet}, girth, M_{\rm jet}, N_{\rm const}, Q^{\rm ch}_{\kappa})
\label{Eq:X}
\end{equation}
A single hidden layer is used with eight
fully connected neurons using a rectified linear unit (ReLU) activation, and an output layer with one neuron employing a sigmoid activation.
\vspace{-0.1cm}
\[
\text{Sigmoid}(x) = \frac{1}{1 + e^{-x}}
\]

The sigmoid output defines the classifier response MLP Score $\in[0,1]$,
interpreted as a measure of the quark-like of the jet as shown in Fig~\ref{fig:MLPScoreQvsG}. This quantity is interpreted as a measure of the quark-likeness of the jet. For a given working point, jets with  MLP Score $<0.5$ are classified as quark-like, while those with MLP Score $> 0.5$ are classified as gluon-like.
This binary labeling is used to construct quark- and gluon-enriched in \gammaJet\ sample and to evaluate classifier performance in terms of efficiencies and misidentification rates. Finally, the performance of the MLP is evaluated using standard metric known as the Receiver Operating Characteristic (ROC) curves discussed in the results Sec~\ref{sec:Results}. \\

\begin{figure*}[t]
  \centering
  \includegraphics[width=0.9\textwidth]{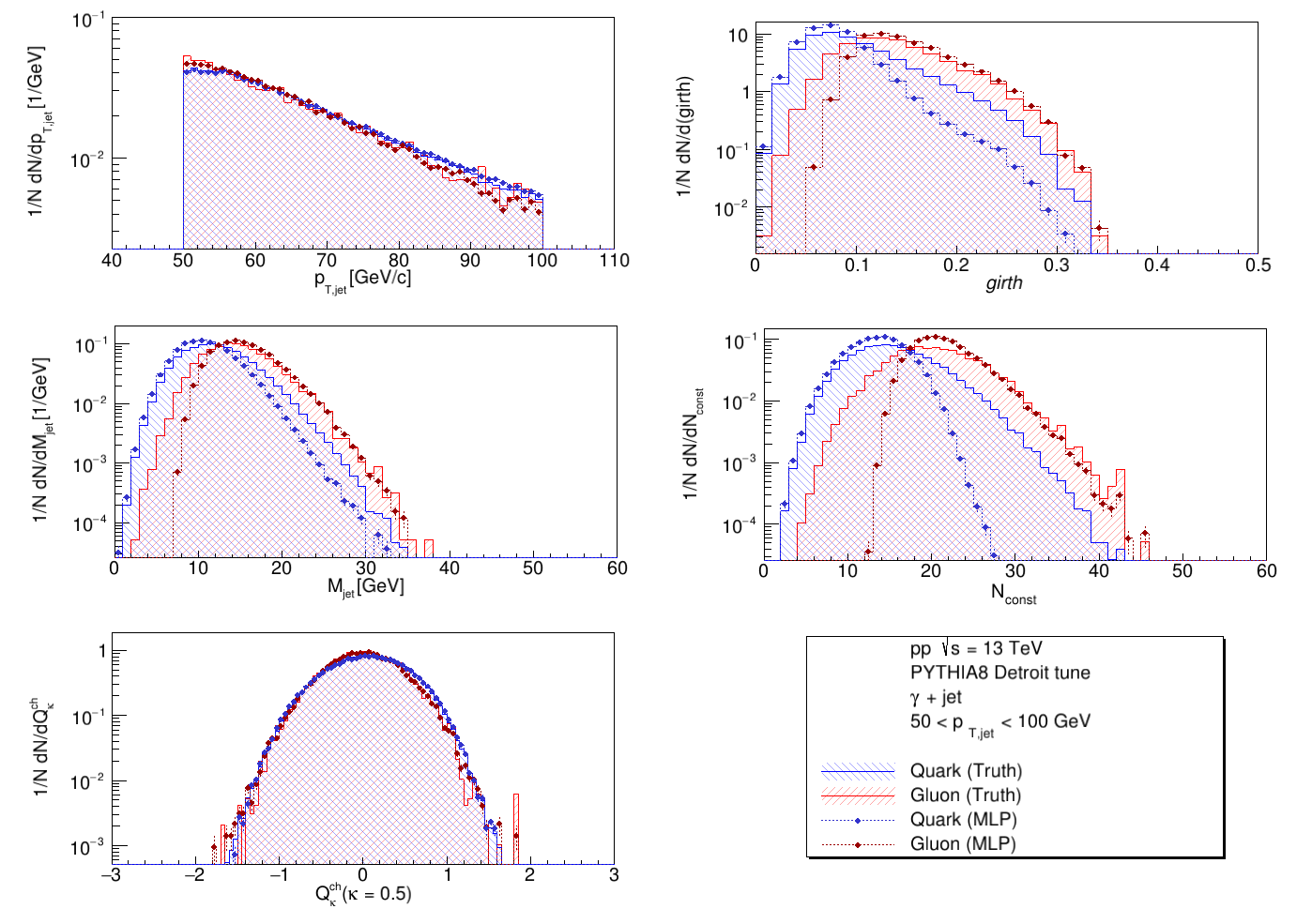}
  \caption{Using the MLP model: \JetPt, \girth, \JetMass, \JetNConst, and \JetCharge distributions are shown for \gammaJet\ with 50 $<$ \JetPt $<$ 100 \gevc . Here, the $y$-axis is normalized by the total number of jets, $N$. The blue hashed region and blue marker represent the true quark and predicted quark-like jet (from QCD Compton process); the same for red color represents gluon jet (from the annihilation process).   }
  \label{fig:MLPJet50to100}
\end{figure*}

\begin{figure*}
    \centering
    \includegraphics[width=0.9\textwidth]{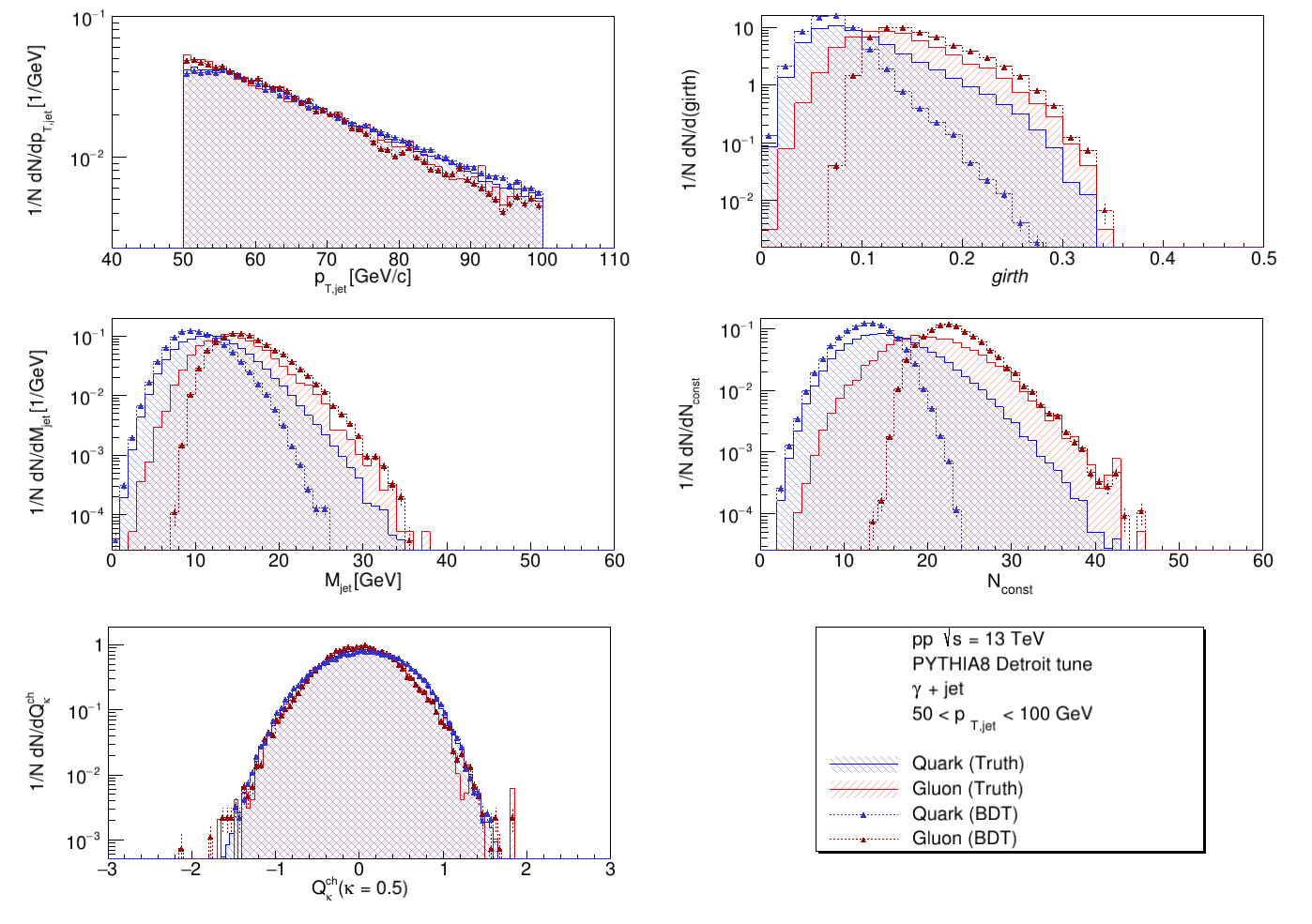}
    \caption{Same for the BDT model.}
    \label{fig:BDT_Jet50to100}
\end{figure*}

\subsubsection{Boosted Decision Tree (BDT)}

The BDT classifier is employed as a complementary and more interpretable multivariate approach to jet substructure observable based process discrimination. The same inputs, as defined in ~\ref{Eq:X} are used in the BDT. The Toolkit for Multivariate Analysis (TMVA)~\cite{Hoecker:2007} within the ROOT~\cite{Brun:1997pa} framework is used and trained in a supervised manner. As in the case of the MLP, the trained BDT classifier is subsequently applied to \gammaJet\ events to discriminate between QCD Compton and quark–antiquark annihilation  processes, based on the flavor of the recoiling jet.\\

The BDT is constructed as an ensemble of decision trees combined using an adaptive boosting algorithm as implemented in TMVA. Each individual tree performs a sequence of binary splits on the input observables, and the final classifier response is obtained as a weighted sum of the tree outputs. The training is performed on statistically independent samples, with separate validation samples used to monitor performance and to reduce the impact of overtraining.\\

In the BDT classifier, gradient boosting is employed with an ensemble of 1000 decision trees, each with
a maximum depth of five. To improve stability and reduce overtraining, a learning rate (shrinkage) of 0.05 is used together with bagging, where each tree is trained on a random 80\% subset of the training sample.
Node splits are optimized using the Gini index as the separation criterion, and cost--complexity pruning is applied to further regularize the trees. These settings provide a balance between classification performance and robustness; the ROC is discussed in the result~\ref{sec:Results}.

\begin{figure}
    \centering
    \includegraphics[width=0.5\textwidth]{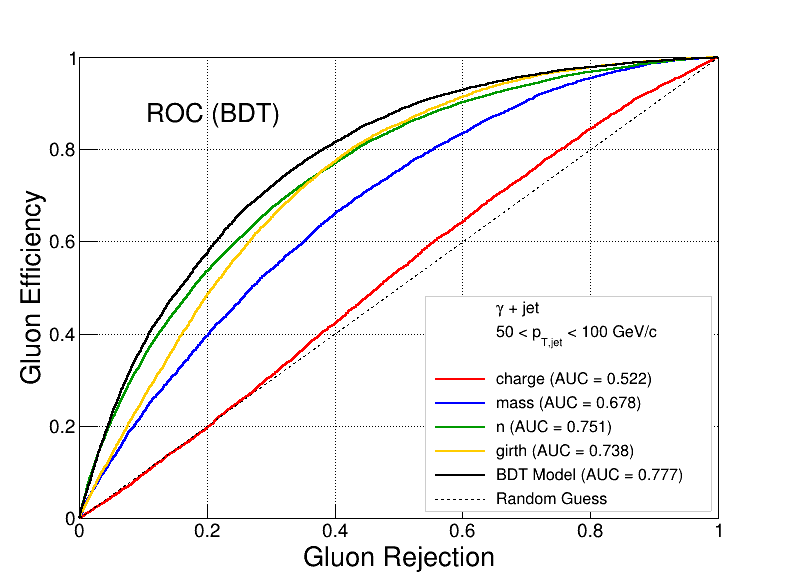}
        \includegraphics[width=0.5\textwidth]{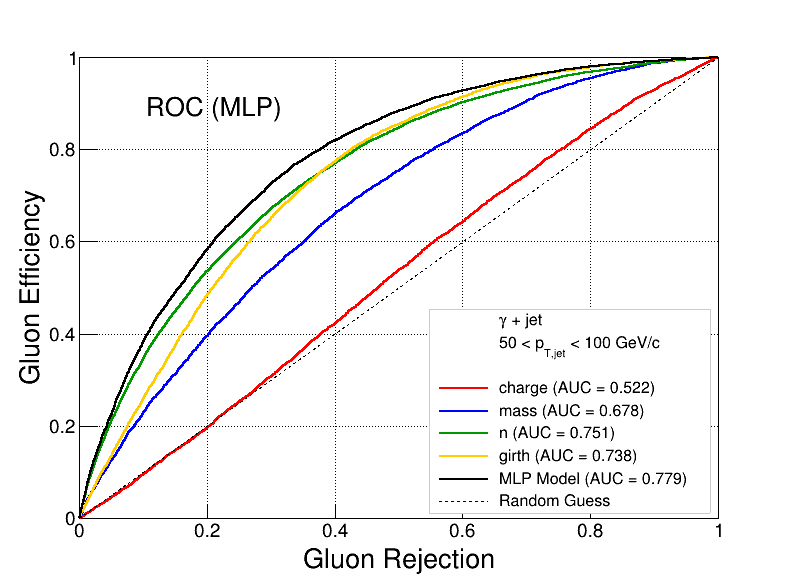}
    \caption{The ROC for the MLP and BDT classifiers. Contributions for different jet observables are also shown.}
    \label{fig:ROC}
\end{figure}

\begin{figure}
    \centering
    \includegraphics[width=0.5\textwidth]{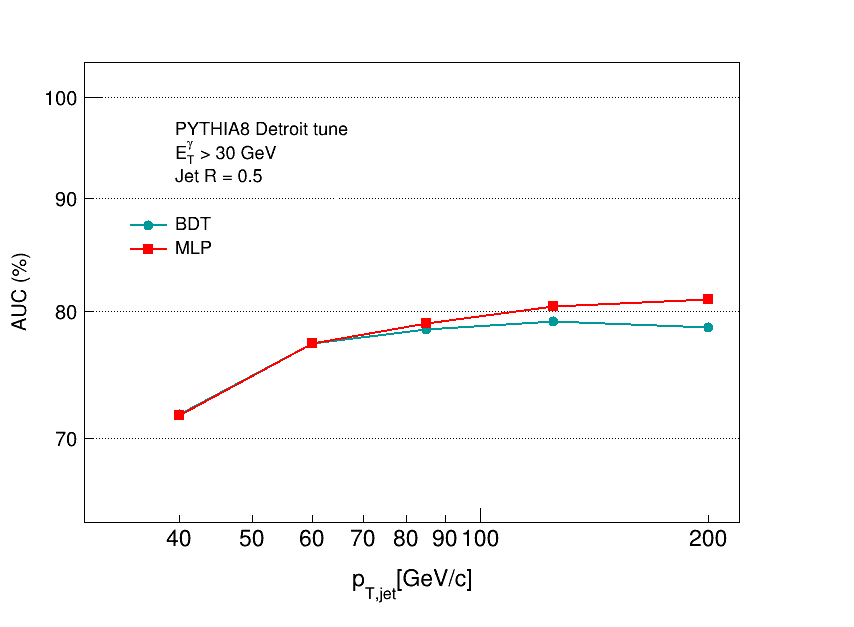}
    \caption{AUC vs \JetPt\ for the BDT and MLP.}
    \label{fig:AUCvsJetPt}
\end{figure}

\section{Results and discussion}
\label{sec:Results}

Figure~\ref{fig:MLPJet50to100} shows the distributions of \JetPt, \girth, \JetMass, \JetNConst, and \JetCharge observables
for truth-labeled and MLP-selected quark- and gluon-initiated jets in
$\gamma+\mathrm{jet}$ events using the MLP score show in Fig~\ref{fig:MLPScoreQvsG}. These jet substructure observables are calculated for 50 $<$ \JetPt $<$ 100 \gevc and jet R=0.5. 
For the girth, jet mass, and jet multiplicity observables, the quark-like MLP-selected samples populate the lower-value region characteristic of truth-labeled quark jets, while the gluon-like samples preferentially occupy the higher-value region associated with gluon jets. The \JetCharge\ shows no noticeable discriminative power. However, the \JetPt\ distributions remain consistent between truth and MLP-selected samples, demonstrating that the classifier does not introduce a kinematic bias. The substantial overlap between quark- and gluon-like distributions persists, reflecting the intrinsic limitations of quark--gluon discrimination imposed by QCD radiation and hadronization.\\

The same observables are also compared for the BDT model in Fig~\ref{fig:BDT_Jet50to100}. In this case, we use two different BDT scores to discriminate quark-like and gluon-like jets as shown in Fig~\ref{fig:BDTScoreQvsG}. The same observation is seen in this case too.\\

The variable-importance ranking obtained from the training shows that observables sensitive to soft and wide-angle radiation, in particular the jet multiplicity and jet girth, dominate the discrimination between quark- and gluon-initiated jets. The jet mass provides a secondary contribution, while the jet charge exhibits negligible importance. This ordering mirrors the behavior seen in the single-variable ROC curves and reflects the larger color factor associated with gluon radiation. The consistency between the variable ranking and the known QCD expectations demonstrates that the multivariate discrimination is driven by physically meaningful features rather than by algorithmic artifacts.
\\


These results demonstrate that both MLP and BDT models effectively and equally enriches quark- and gluon-dominated jet samples by exploiting correlations among substructure observables, while remaining consistent with established QCD expectations.\\

To quantify the correlation between signal efficiency and background rejection as a function of the classifier threshold (BDT and MLP scores), 
ROC curves are constructed for both models. Figure~\ref{fig:ROC} shows the gluon efficiency as a function of gluon rejection using individual jet observables as well as multivariate classifiers based on the BDT and MLP. Here the gluon efficiency ($\epsilon_{g}$) is defined as the ratio of number of true gluon jets correctly tagged as gluon over the total true gluon jets. The Gluon rejection is $R_{g} = 1 - \epsilon_{q}$; where $\epsilon_{q}$ is the number of quark jets misidentified as gluon over the total quark jets.

Among the single observables, the jet multiplicity and jet girth exhibit the strongest discriminating power, whereas the jet charge provides negligible separation. This behavior is consistent with the larger color factor associated with gluon radiation in perturbative QCD, as implemented in the PYTHIA parton-shower framework, leading to higher particle multiplicity and broader energy flow in gluon-initiated compared to quark-initiated jets. In the experimental condition, it would be even challenging due to detector effect and pileup contamination.\\

To study the kinematic dependence of the discrimination performance, the area under the curve (AUC) is evaluated as
\begin{equation}
    AUC = \int_{0}^{1} \epsilon_{g} (R_{g}) dR_{g}.
\end{equation}

The $AUC$ as a function of \JetPt\ for both the BDT and MLP classifiers is shown in Fig.~\ref{fig:AUCvsJetPt}. The discrimination power increases from low to intermediate jet $p_T$, reflecting the improved resolution of jet substructure as the available radiation phase space grows. At higher $p_T$, the performance saturates, indicating that additional radiation does not lead to further separation between quark- and gluon-initiated jets. The close agreement between the BDT and MLP results across the full \JetPt\ range demonstrates that this saturation is driven by QCD dynamics rather than by the expressive power of the classification algorithms.\\

Establishing this consistent classical performance, it is important to test quark--gluon jet discrimination across LHC, RHIC, and EIC relevant kinematic regimes in order to validate a robust multivariate baseline. Such a baseline may also serve as a useful reference for exploring alternative learning strategies in the future, including quantum-inspired approaches, in regimes where classical methods appear to saturate.


\section{summary and Outlook}
\label{sec:SummaryOutlook}

We study the discrimination of QCD Compton and quark--antiquark annihilation processes in $\gamma+$jet events in \pp\ collisions at $\sqrt{s}=13$ TeV, using multivariate classifiers based on BDT and MLP, with events simulated using the PYTHIA8 Detroit tune. The classifiers are trained on dijet quark- and gluon-initiated samples, their responses are interpreted as measures of quark- or gluon-like when applied to \gammaJet\ events, enabling an indirect separation of the underlying production mechanisms.\\

Among the individual jet observables, the jet multiplicity and jet girth exhibit the strongest discriminating power. The jet mass provides a complementary but weaker contribution. The jet charge provides negligible separation. The BDT and MLP based analyses further enhance the discrimination by exploiting correlations among jet substructure variables, yielding quark- and gluon-enriched jet samples that retain physically meaningful characteristics. The observed separation originates from well-understood differences in QCD radiation patterns rather than from kinematic biases. Here the substantial overlap between quark and gluon jets reflects an intrinsic limitation imposed by QCD dynamics, consistent with theoretical expectations.\\

The close agreement between the BDT and MLP results indicates that the dominant discriminating information is already captured by the selected set of jet observables, with the observed performance saturation driven by physics rather than algorithmic limitations. In future, it would be important to incorporate the higher-dimensional jet substructure information to correlate QCD effects and to implement the advanced machine learning approach~\cite{Gong:2022lye,Datta:2019ndh,Almeida:2015jua}. These approaches are particularly well suited to precision jet studies at the LHC and the future EIC, where controlled kinematics and reduced backgrounds enable detailed exploration of correlated QCD dynamics.\\

 
\begin{acknowledgments}

\textbf{Acknowledgments—} 
MS thanks IISER Tirupati for hospitality during the course of this work. NRS acknowledges support from the IISER Tirupati startup grant.
\end{acknowledgments}

\section{Appendix}
\appendix
\section{Quark- vs Gluon-initiated jet comparison from Dijet and \gammaJet\ sample}
\label{Appn:DiGammaJet}
\begin{figure*}
    \centering
    \includegraphics[width=0.8\textwidth]{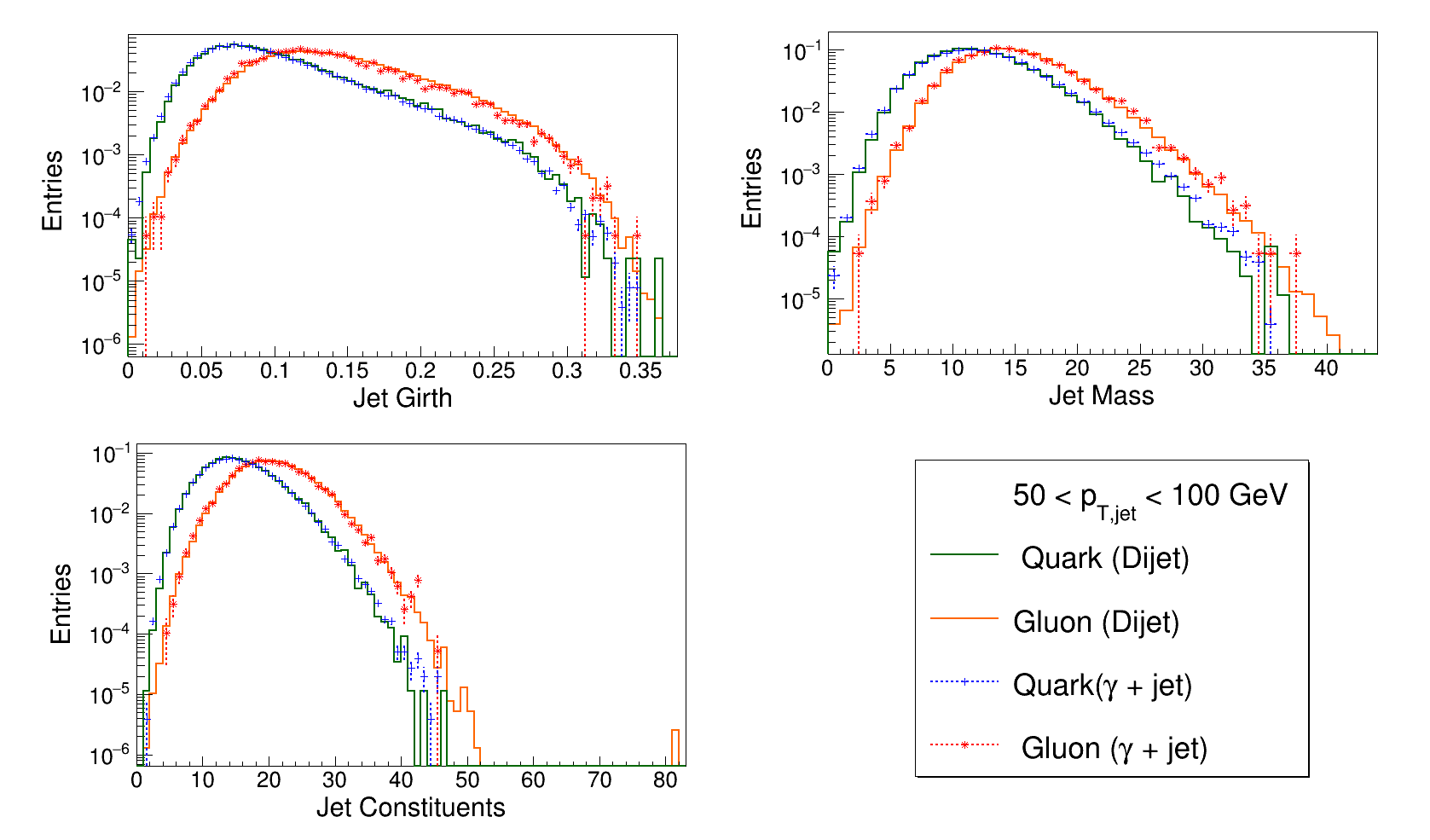}
    \caption{Quark vs Gluon-initiated jet comparison from Dijet and \gammaJet\ sample: \girth, \JetMass, and \JetNConst. Green line and blue makers represent quark-jets; Orange line and red markers represent gluon-jets.}
    \label{fig:DiJetGammaJet}
\end{figure*}

 Figure~\ref{fig:DiJetGammaJet} shows a comparison of jet girth and jet multiplicity for quark- and gluon-initiated jets in dijet and \gammaJet\ events within the same jet \JetPt\ interval, as simulated with the PYTHIA8 Detroit tune. For both jet flavors, the distributions are found to be similar across the two event topologies, indicating that the jet substructure features learned from dijet events are transferable to \gammaJet\ recoil jets within uncertainties.

\bibliographystyle{apsrev4-2}


\bibliography{references}

\end{document}